\documentclass[aps,prx,twocolumn,showpacs,showkeys,superscriptaddress,floatfix]{revtex4-1}

\usepackage{graphicx}
\usepackage{bm}
\usepackage{hyperref}
\usepackage{grffile} 
\usepackage{amsmath}
\usepackage{natbib}
\usepackage{color}
\usepackage{mathtools}
\usepackage{float}
\usepackage{graphics}
\usepackage{epsfig}
\usepackage{bbold}
\usepackage{verbatim}
\usepackage{color,hyperref}
\usepackage[dvipsnames]{xcolor}
\usepackage{natbib}
\definecolor{darkblue}{rgb}{0.0,0.0,0.0}
\hypersetup{colorlinks,breaklinks,
           linkcolor=darkblue,urlcolor=darkblue,
           anchorcolor=darkblue,citecolor=darkblue}
\newcommand{\EQ}[1]{Eq.~(\ref{eq:#1})}
\newcommand{\EQS}[2]{Eqs.~(\ref{eq:#1}) and (\ref{eq:#2})}
\newcommand{\EQSrange}[2]{Eqs.~(\ref{eq:#1} - \ref{eq:#2})}
\newcommand{\FIG}[1]{Fig.~\ref{fig:#1}}

\begin{document}

\title{Nonreversible Markov chain Monte Carlo algorithm for efficient generation of Self-Avoiding Walks}
\author{Hanqing Zhao}
\affiliation{Department of Physics, University of Virginia, Charlottesville, VA 22904, USA}
\author{Marija Vucelja}
\email{mvucelja@virginia.edu}
\affiliation{Department of Physics, University of Virginia, Charlottesville, VA 22904, USA}
\date{\today}


\begin{abstract}
We introduce an efficient nonreversible Markov chain Monte Carlo algorithm to generate self-avoiding walks with a variable endpoint. In two dimensions, the new algorithm slightly outperforms the \emph{two-move nonreversible Berretti-Sokal algorithm} introduced by H.~Hu, X.~Chen, and Y.~Deng  in~\cite{old}, while for three-dimensional walks, it is 3--5 times faster. The new algorithm introduces nonreversible Markov chains that obey global balance and allow for three types of elementary moves on the existing self-avoiding walk: shorten, extend or alter conformation without changing the length of the walk. 
\end{abstract}

\maketitle

\section{Introduction}
A Self-Avoiding Walk (SAW) is defined as a contiguous sequence of moves on a lattice that does not cross itself; it does not visit the same point more than once. SAWs are fractals with fractal dimension $4/3$ in two dimensions, close to $5/3$ in three dimensions, and $2$ in dimensions above four ~\cite{Havlin_1982,PhysRevA.26.1728}. In particular two-dimensional SAWs are conjectured to be the scaling limit of a family of random planar curves given by the Schramm-Loewner evolution with parameter $\kappa = 8/3$~\cite{2002math......4277L}. Since their introduction, SAWs have been used to model linear polymers~\cite{53Flory,mc1,mcmc}. They are essential for studies of polymer enumeration where scaling theory, numerical approaches, and field theory are too hard to analyse~\cite{poly1,poly2}.
SAWs are also used in the numerical studies of finite-scaling~\cite{ising1} and two-point functions~\cite{ising2} of Ising model and $n-$vector spin model~\cite{nspin}.
Analytical results on SAWs are scarce, and generating long SAWs is computationally complex.

Typically one uses Monte Carlo approaches~\cite{97Sokal, NewmanMC} to generate SAWs numerically. Many previous Markov chain Monte Carlo (MCMC) algorithms have been designed to efficiently produce different kinds of SAWs by manipulating potential constructions that can be executed on a walk to increase, decrease its length, or change its conformation. For example, the pivot algorithm samples fixed-length SAWs -- it alters the walk's shape without changing its length~\cite{pivot}. While the Berretti-Sokal algorithm and BFACF algorithm contain length-changing moves and can generate walks with varying lengths~\cite{bsalgorithm, BFACF}.

The above described MCMC algorithms satisfy the detailed balance condition  -- which states that the weighted probabilities of transitions between states are equal. In other words, these algorithms use reversible Markov chains. The reversibility introduces a diffusion-like behavior in the space of states. In recent years, there has been progress in designing nonreversible Markov chains that converge to the correct target distribution. Such chains due to "inertia" reduce the diffusive behavior, sometimes leading to better convergence and mixing properties compared to the reversible chains~\cite{DHN97,CLP00,turitsyn2011irreversible,LiftingVucelja,2013SakaiHukushima,bierkens2017,joris,17KapferKrauth}.

As for SAW, H.~Hu, X.~Chen, and Y.~Deng modified the Berretti-Sokal algorithm to allow for nonreversible Markov chains~\cite{old}. This modification yields about a ten times faster convergence than the original Berretti-Sokal algorithm in two dimensions and is even more superior in higher dimensions. Both the original and the modified Berretti-Sokal algorithm have two elementary moves -- to shorten or extend the SAW.  Building upon these algorithms, we add another move -- to alter the conformation of SAW and introduce a three-move nonreversible MCMC technique to create SAWs. We discuss the advantages of this approach and compare the two nonreversible algorithms. The three types of moves correspond to three types of "atmospheres"; therefore, we start below by defining an atmosphere. 

\section{The atmospheres}
The algorithms creating SAWs usually manipulate different kinds of proposed moves, often referred to as \emph{atmospheres}~\cite{atm0,atm1,atm2,endpointatm}. Atmospheres can be described as potential constructions that can be executed on a given walk to increase or decrease the current length or change the conformation. When generating SAWs, the algorithm usually performs moves on either endpoint atmospheres or generalized atmospheres where \emph{positive} and \emph{negative} atmospheres are generally defined as ways of adding or removing a fixed number of edges to the current walk. In contrast, \emph{neutral} moves are ways of altering the walk's shape without changing its length. For instance, the pivot algorithm, which only acts on neutral atmospheres, can be used to sample fixed-length walks~\cite{pivot}. In contrast, the Berretti-Sokal algorithm and BFACF algorithm contain length-changing atmospheric moves and can generate walks of different lengths~\cite{bsalgorithm, BFACF}.

Suppose $s$ is the current SAW starting from the origin with length $|s|$ and its last vertex is $v$. The positive endpoint atmospheres are the lattice edges incident with the last vertex, which can be occupied to extend the length by one. The negative endpoint atmosphere is just the last occupied edge since removing it can extract the length by one. The neutral endpoint atmospheres are edges that can be occupied by changing the direction of the vertex $v$. For any SAW with a non-zero length, the number of negative endpoint atmospheres is one. If the SAW has zero length, the number of negative endpoint atmospheres is set to zero, as the length can not be further reduced. 

\begin{figure}[htp]
\includegraphics[width=0.5\columnwidth]{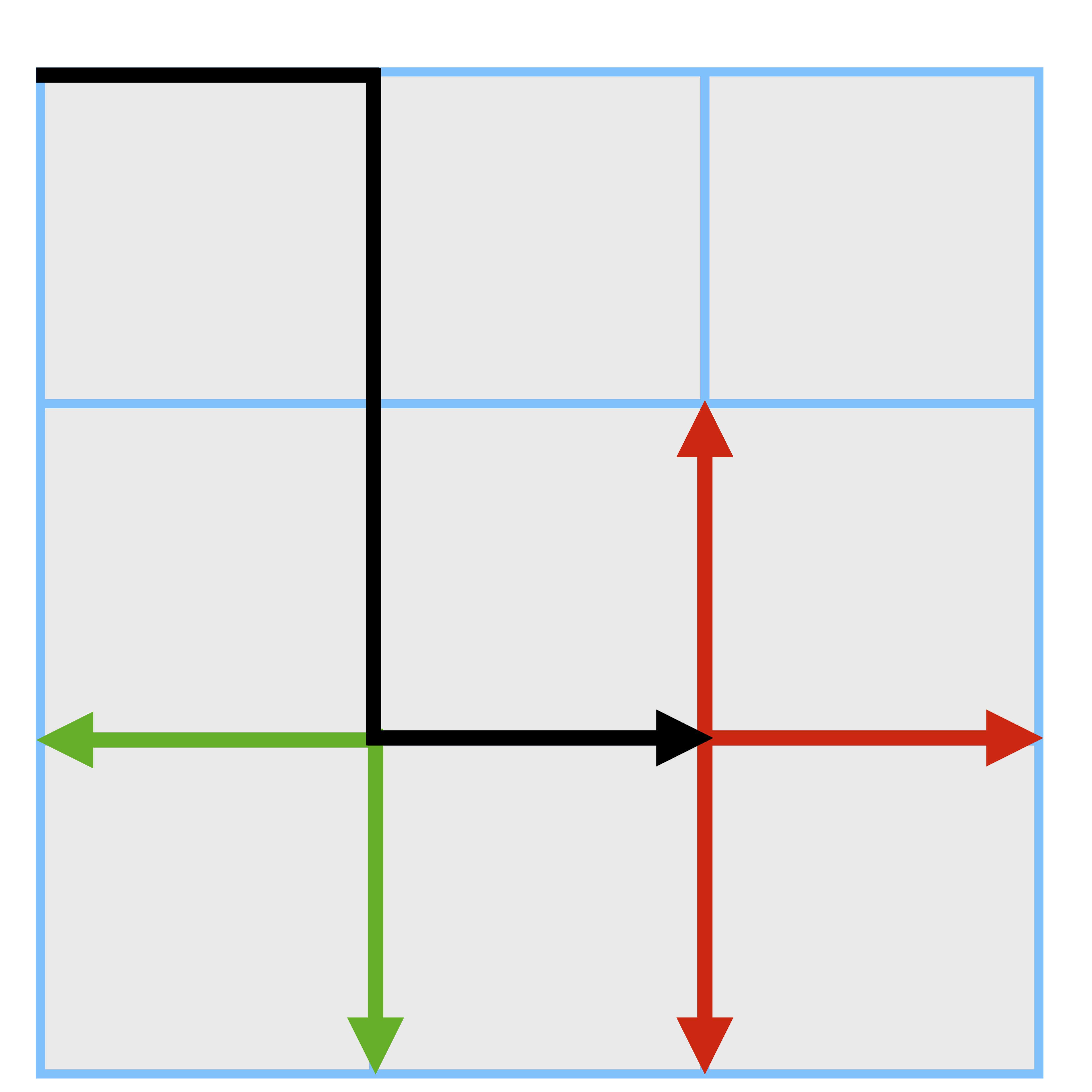}
\caption{\label{fig:atm-v01.jpg}The endpoint atmospheres on a self-avoiding walk of length $|s| = 4$. For this self-avoiding walk, there are three positive ending atmospheres (red arrows) and one endpoint atmosphere, which is the last occupied edge (black arrow), and the number of neutral endpoint atmospheres is two (green arrows).}
\end{figure}

\FIG{atm-v01.jpg} shows a SAW with a length equal to four. In this example, three unoccupied edges are incident with the last vertex; they are shown in red on the graph, making three positive ending atmospheres. As we see from the last occupied edge (black arrow), there is just one negative endpoint atmosphere. There are two neutral endpoint atmospheres, and the corresponding edges are displayed with green arrows.

\begin{figure}[htp]
\includegraphics[width=\columnwidth]{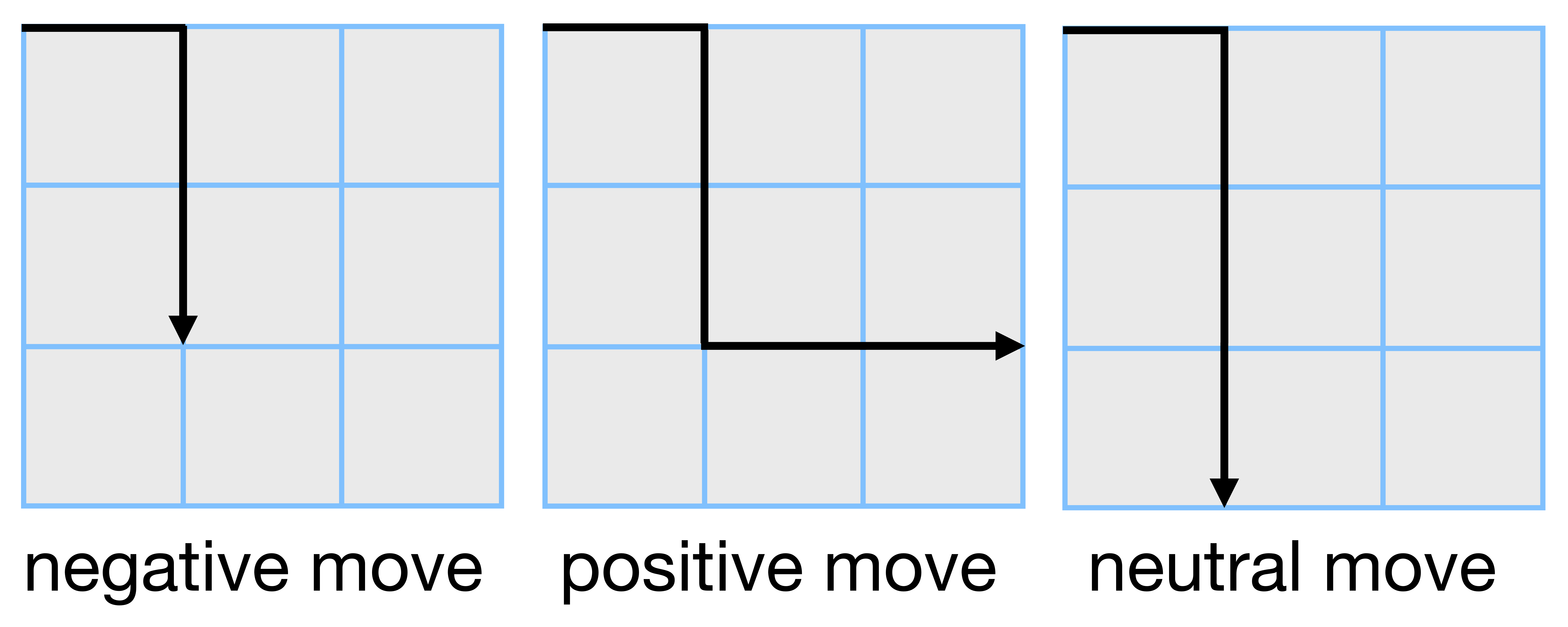}
\caption{\label{fig:fig-negative-positive-neutral-move-v01.jpg}Possible self-avoiding walks after executing one move on the self-avoiding walk shown in~\FIG{atm-v01.jpg}.}
\end{figure}

Three types of elementary moves in the algorithm executing the endpoint atmospheres correspond to the three kinds of endpoint atmospheres. Here we call a \emph{positive move} the one to be performed on a positive endpoint atmosphere, resulting in occupying one empty edge incident with the last vertex. Similarly, a \emph{negative move} implies executing on the negative endpoint atmosphere, that is, deleting the last occupied edge. Finally, the \emph{neutral move} is changing the direction of the last occupied edge. The three kinds of moves' for the SAW in~\FIG{atm-v01.jpg} are illustrated in~\FIG{fig-negative-positive-neutral-move-v01.jpg}.

\section{The Berretti-Sokal algorithm}
The balance condition is one of the most important factors in designing an MCMC algorithm since it ensures that the Markov chain will converge to a target distribution. The balance condition for most MCMCs is the so-called Detailed Balance Condition (DBC)
\begin{align}
\label{eq:DBC}
    P_{ij}\pi_j=P_{ji}\pi_i, \quad \forall i,j \in \Omega,
\end{align}
where $P_{ij}$ is the transition probability from state $j$ to state $i$, $\Omega$ is the space of states, and $\pi$ is the stationary distribution, see e.g.~\cite{2009Levinbook,LiftingVucelja}. Detailed balance is a local condition and thus easy to implement. However, for a Markov chain to asymptotically converge to a stationary distribution $\pi$, all we need is a weaker condition -- the Global Balance Condition (GBC): 
\begin{align}
\label{eq:GBCabstract}
    \sum _{j\in \Omega} P_{ij}\pi_j=\sum _{j\in \Omega} P_{ji}\pi_i, \quad \forall i \in \Omega,  
\end{align}
where $\Omega$ is a space of states. The GBC physically means that the total probability influx at a state equals the total probability efflux from that state~\cite{turitsyn2011irreversible,97Sokal}.

Note that the probability distribution of a SAW of length $|s|$ is 
\begin{align}
    \pi = x^{|s|}
\end{align}
where $x$ is the weight of a unit step. This is what we want the Markov chain target distribution to be. 

One of the most famous reversible MCMC algorithms that manipulate the endpoint atmospheres is the Berretti-Sokal algorithm~\cite{bsalgorithm}. The Berretti-Sokal algorithm only considers the positive and negative endpoint atmospheres and thus has two elementary moves: the increasing and the decreasing move. In this paper, we are using a Metropolis-Hastings style~\cite{MRRTT53, H70} implementation of the Berretti-Sokal algorithm. It works as follows:
\begin{itemize}
    \item[(i)] Suppose the current length of a SAW is given by $N$. With equal probability, the algorithm chooses the increasing move or the decreasing move.
    \item[(ii)] If the increasing move is selected, with probability $P_+$ one of the empty edges incident with $v_N$, the last vertex, will be occupied randomly when this leads to a valid SAW of $N+1$ steps. Similarly, for the decreasing move, the last occupied edge is deleted with probability $P_-$.
    The two probabilities are given by 
    \begin{align}
        \label{eq:Pplus}
       &P_+=\min\{1, x(z-1)\},
       \\
       \label{eq:Pminus}
       &P_-=\min\left\{1,\frac{1}{x(z-1)}\right\},
    \end{align}
    where $z$ is the coordination number of the system, i.e. the number of lattice points neighboring a vertex on the lattice.
\end{itemize}

Special attention is needed for the "null" walk, $|s| = 0$, in such case only an increasing mode is allowed and the number of empty edges is $z$, rather than $z-1$. For simplicity we permanently set $P_+=\min\{1, x(z-1)\}$.

To prove that the DBC holds in the Berretti-Sokal algorithm, let us for example consider the case where $x(z-1)<1$. From~\EQS{Pplus}{Pminus} we conclude that the choice implies $P_+<1$ and $P_-=1$. Thus we have $x^{|s|}P_+(z-1)^{-1}=x^{|s+1|} = x^{|s+1|}P_-$, which satisfies the DBC, given in~\EQ{DBC}. The proof is analogous in the case $x(z-1) >1$. 

\section{Nonreversible Berretti-Sokal Algorithms}
One possible way to set up a nonreversible algorithm is to increase the phase space by introducing replicas~\cite{turitsyn2011irreversible,old,LiftingVucelja} and work on the extended space with nonzero probability fluxes. Here we follow an analogous approach. As mentioned above, there has been a successful \emph{two-move nonreversible Berretti-Sokal algorithm}~\cite{old}. The authors achieved an important improvement in the speed of the algorithm. The speedup is about tenfold in two-dimensional systems and is even more pronounced in higher-dimensional systems. They set up two modes in the algorithm, which we call the increasing mode and the decreasing mode. 

\subsection{Three-move Nonreversible Berretti-Sokal Algorithm}
\begin{figure}[htp]
\includegraphics[width=\columnwidth]{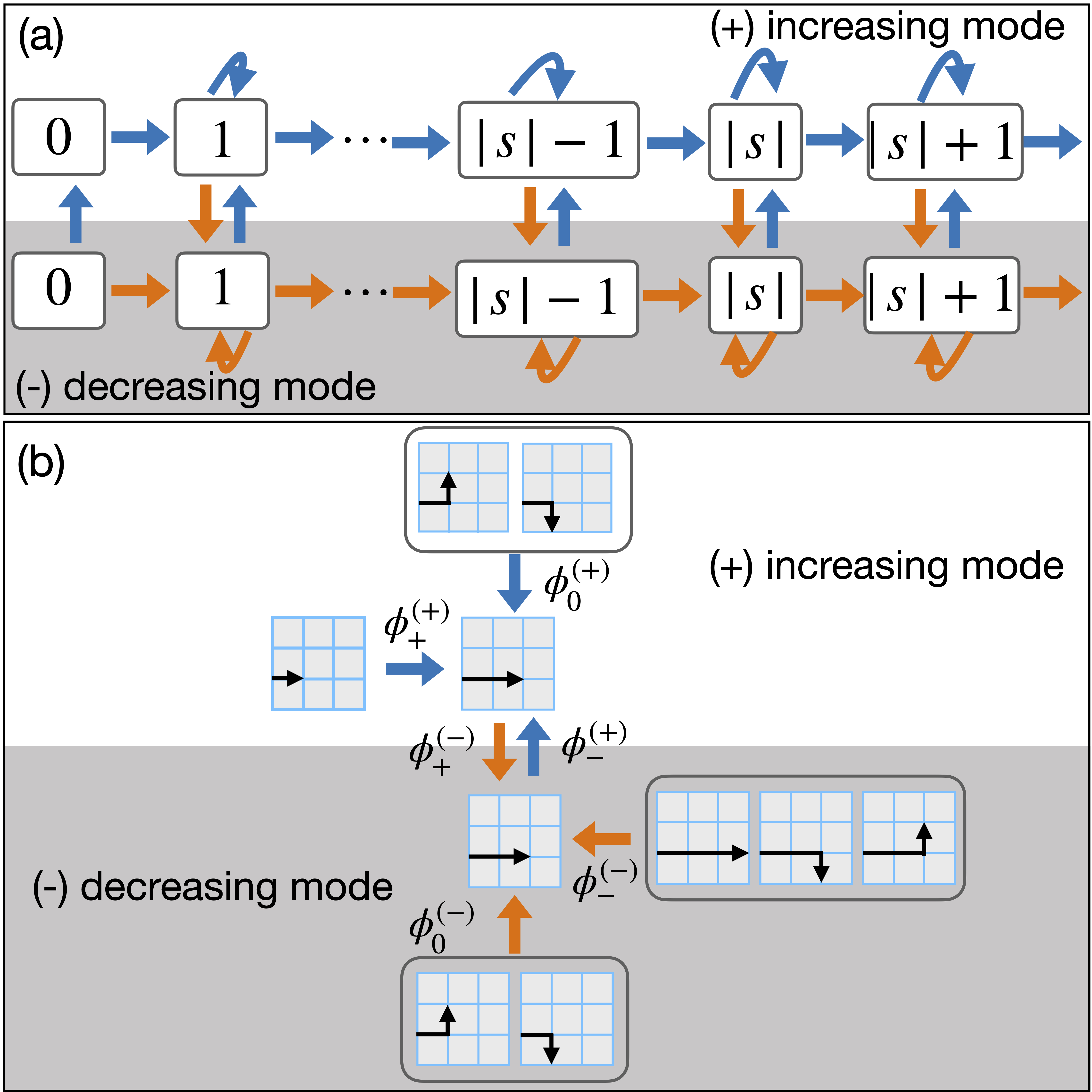}
\caption{\label{fig:figure-example-full.jpg}{\bf (a)} Diagram of probability flows in the \emph{three-move nonreversible Berretti-Sokal algorithm}. Each rectangle specifies a SAW of length $|s|$. Each realization of the algorithm is different because of the neutral moves, allowing to alter the configuration of the walk. The top row represents the \emph{increasing mode} in which the algorithm can produce either a positive or neutral move, while the bottom row represents the \emph{decreasing mode} where the algorithm produces either negative or neutral moves. The circular arrow represents the execution of a neutral move, leading to a SAW with the same length but a different shape as the last occupied edge's direction is changed. The 'null' walk, $|s| = 0$, requires special attention; in this case, we do not allow neutral and decreasing moves.
{\bf(b)} Example of the incoming fluxes for SAW of length $|s| = 2$ in $2D$ on a square lattice.}
\end{figure}
The new algorithm has a third type of move -- besides shortening and extending the SAW, we also allow the SAW to change its conformation. Namely, in the increasing mode, the algorithm can perform either an increasing move or a neutral move; in this mode, the decreasing move is not allowed. Analogously, in the decreasing mode, the algorithm will only execute either a decreasing move or a neutral move. A diagram describing the algorithm is shown in ~\FIG{figure-example-full.jpg}. It works as follows:
\begin{itemize}
    \item[i)]In the increasing mode, with equal probability, perform either the \emph{positive move} or the \emph{neutral move}. For the \emph {positive move}, the algorithm will randomly occupy one of the empty edges incident to the last vertex with probability $P_+$. While for the \emph{neutral move}, the algorithm will change the direction of its last occupied edge randomly. If the chosen move does not lead to a valid SAW, the algorithm will change to the decreasing mode.
    \item[ii)]In the decreasing mode, with equal probability, perform either the \emph{negative move} or the \emph{neutral move}. For the negative move, the algorithm will delete the last occupied edge with probability $P_-$. For the \emph{neutral move}, the algorithm will change the direction of its last occupied edge randomly. If the chosen move does not lead to a valid SAW, the algorithm will change into the increasing mode.
    \item[iii)]When the length is 0, the algorithm will be changed into the increasing mode, and a \emph{positive move} will be performed.
\end{itemize}
Therefore, in each step, the algorithm will either execute one of the elementary moves successfully or change to the other mode. The global balance condition implies that the total influx probability flow equals the efflux probability flow; that is, we have 
\begin{align}
\label{eq:GBC}
\phi^{(\pm)}_{\pm}+\phi^{(\pm)}_{0}+\phi^{(\pm)}_{\mp}=x^{|s|},
\end{align}
where $x^{|s|}$ is the distribution of SAWs of length $|s|$ and $\phi-$s describe the incoming probability fluxes, where the superscript denotes the mode and the subscripts denote the move. The three terms on LHS are the incoming flow of executing a $\pm$ move in mode $(\pm)$, $\phi^{(\pm)}_{\pm}$, the incoming flow of executing one neutral move in mode $(\pm)$, $\phi^{(\pm)}_0$, and the incoming flow from switching the mode from $(\mp)$ to $(\pm)$, $\phi^{(\pm)}_{\mp}$. To clarify the third term in the LHS by example: $\phi^{(+)}_-$ is the incoming flux from switching from $(-)$ mode to the $(+)$ mode. 
Let us show that global balance condition holds for the \emph{increasing mode} when $x(z-1)<1$. Proofs for the other cases follow analogously. In this case the three fluxes are: 
\begin{itemize}
    \item The incoming flux from a positive move is
    \begin{align}
        \label{eq:incoming-flow-from-plus}
        \phi^{(+)}_{+}&=x^{|s|-1}P_+\frac{1}{2(z-1)}=\frac{x^{|s|}}{2}, 
    \end{align}
    where in the second equality we used~\EQ{Pplus}. The factor $1/2$ is the result of selecting either a positive move or a neutral move and the term $(z-1)^{-1}$ is from occupying one of the $z-1$ empty edges incident to the last vertex.
    \item The incoming flux from a neutral move is 
    \begin{align}
        \label{eq:incoming-flow-from-neutral}
        \phi^{(+)}_{0}=\frac{x^{|s|}z''}{2(z-1)},
    \end{align} 
    where $z''$ is the number of possible edges which will lead to a valid SAW for the last occupied edge when changing its direction.
    \item The incoming flux from the decreasing mode, $\phi^{(+)}_{-}$, since $P_-=1$, as we assume that $x(z-1)<1$, the only possible reason of changing from another mode is that when the last occupied changes it direction, it does not lead to a valid SAW, thus
    \begin{align}
        \label{eq:incoming-flow-from-minus}
        \phi^{(+)}_{-} =\frac{1}{2}x^{|s|}\left(1-\frac{z''}{z-1}\right).
    \end{align}
\end{itemize}
Summing over the incoming flows, given in~\EQSrange{incoming-flow-from-plus}{incoming-flow-from-minus}, we verify that the global balance condition,~\EQ{GBC}, holds. Note that we do not assume that a particular SAW configuration of length $|s|$ is achieved with the same frequency in the increasing and the decreasing mode -- it comes out as a corollary of the global balance condition.

\begin{figure}[htp]
\includegraphics[width=\columnwidth]{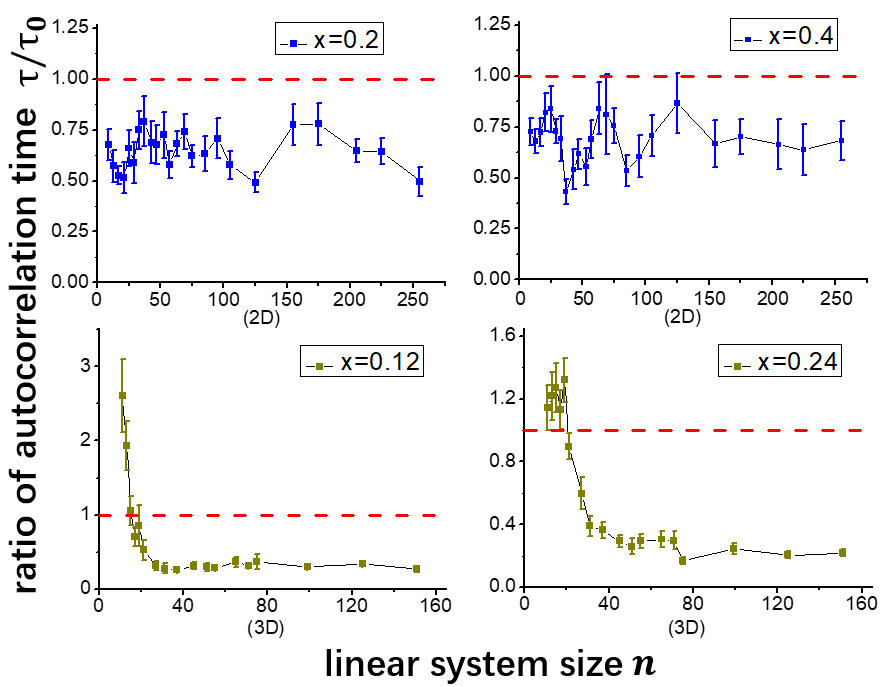}
\caption{\label{fig:newold.jpg}The ratio of integrated autocorrelation times of the \emph{three-move nonreversible Berretti-Sokal algorithm}, $\tau$, and the \emph{two-move nonreversible Berretti-Sokal algorithm}, $\tau_0$, for $2D$ and $3D$ systems as a function of the linear system size $n$. The \emph{three-move nonreversible Berretti-Sokal algorithm}'s performance is slightly better in $2D$ systems while it is $3-5$ times faster in most $3D$ systems.}
\end{figure}
To test the efficiency of the new algorithm, we used the integrated autocorrelation time $\tau$. For a given observable $\mathcal{O}$, it is defined as
\begin{align}
    \tau = \frac{m}{2}\frac{\sigma^{2}_{\overline{\mathcal{O}}}}{\sigma^{2}_{\mathcal{O}}}, 
\end{align}
where $m$ is the number of steps, $\overline{\mathcal{O}}$ is the estimator of the average $\mathcal{O}$, and $\sigma^2$ denotes a variance, c.f.~\cite{autotime}. Here we choose the length of the walk, ${|s|}$, for the observable as it is a common choice for SAWs. We tested the efficiency as a function of the linear system size by generating SAWs in a square lattice with $n\times n$ points and in a cubic lattice with $n \times n\times n$ points. The boundary conditions were fixed. With $\tau_0$ we denote the integrated autocorrelation time of the \emph{two-move nonreversible Berretti-Sokal algorithm} (algorithm from~\cite{old}). The comparison of the two algorithms is on~\FIG{newold.jpg}. 

Note, that there are two different scenarios based on the value of weight of a unit step $x$. For example, for a $2D$ square lattice, when $x=0.4$,  $P_+=1$ and $P_-<1$, while for $x=0.2$, $P_-=1$ and $P_+<1$. To study both scenarios present the results under initial setting where $x=0.2$ and $x=0.4$ in a $2D$ system and correspondingly $x=0.12$ and $x=0.24$ in a $3D$ system. From~\FIG{newold.jpg} we conclude that the ratio of the autocorrelation times for large systems is weakly dependent on the value of $x$.  

In $2D$, the ratio of the autocorrelation time of the new algorithm over the previous one is always less than one, which means that the new algorithm has a slightly better performance. We further tested the new algorithm in a three-dimensional cubic system. The new algorithm tends to have better performance in large systems, and the difference is more significant than the $2D$ situation. When the length of the cube is less than 20, the previous algorithm is more efficient with less autocorrelation time. However, as the system's scale increases, the ratio $\tau/\tau_0$ becomes less than one, and the value is between $0.2$ and $0.3$, indicating that the new algorithm is $3$ to $5$ times faster in these larger $3D$ systems. We have also tested our algorithm in $4D$ and $5D$ systems where 
no general improvements are found compared to the \emph{two-move nonreversible Berretti-Sokal algorithm}. We show the detailed findings in Appendix~\ref{sec:appendix}. The fact that the addition of neutral moves does not improve the efficiency in generating SAWs in $4D$ and $5D$, could be explained by the fact that as dimension gets higher, it will be much more likely for the algorithm to make a successful, positive move, which results in less benefit from adding the neutral move.

To summarize, we have created a new nonreversible algorithm manipulating the endpoint atmospheres to generate SAWs. By introducing all three kinds of endpoint atmospheres' moves, the new algorithm has greater flexibility than the \emph{two-move nonreversible Berretti-Sokal algorithm}, from~\cite{old}. For instance, when occupied lengths surround the endpoint of a given SAW, the algorithm will change into the negative mode since neither a neutral move nor a positive move will lead to a valid SAW. Assume that $P_+<1$, for an algorithm with only positive and negative moves, it will return to the origin and start from the beginning again. On the other hand, with a neutral move, the SAW does not have to start from the origin again. When a neutral move in the negative mode is not possible, the algorithm will change into the positive mode. The addition of neutral moves gives the algorithm greater flexibility in finding valid SAWs. 

\section{Conclusion}
We have created a new nonreversible algorithm manipulating the endpoint atmospheres to generate SAWs. The previous \emph{two-move nonreversible Berretti-Sokal algorithm} has already improved the efficiency greatly as its speed is ten times faster than the original Berretti-Sokal algorithm in $2D$ systems and is even more superior in higher-dimensional systems. By introducing all three kinds of endpoint atmospheres' moves, the \emph{three-move nonreversible Berretti-Sokal algorithm} has greater flexibility and higher efficiency than the two-move algorithm. By comparing the autocorrelation time, the new algorithm is slightly faster in $2D$ systems and is $3$ to $5$ times faster in most $3D$ systems.

The \emph{three-move nonreversible Beretti-Sokal algorithm} is designed to create SAWs with a fixed beginning point and variant ending points. There are also algorithms manipulating general atmospheres instead of endpoint atmospheres. Algorithms like the BFACF algorithm can create SAWs with a fixed beginning and ending point~\cite{BFACF}. Meanwhile, other algorithms generating SAWs like the PERM, GARM, and pivot algorithm have no nonreversible versions yet~\cite{PERM1,PERM2,atm2,pivot}. Previous research has improved the efficiency of PERM algorithm without implementing the nonreversible MCMC techniques~\cite{parrelPERM}. These algorithms might serve as aspects for future research. 

Finally, here we manually found a way with three atmospheres on how to fulfill the global balance. Looking into the future, one might delegate this task to a neural network alike in~\cite{NIPS2017_7099}. Optimizing the transition operator with more than three types of endpoint atmospheres might further increase the efficacy. 

\section{Acknowledgement}
MV and HZ acknowledge discussions with Michael Chertkov, Gia-Wei Chern, Jon Machta, Joris Bierkens, Christoph Andrieu and Chris Sherlock. This material is based upon work supported by the National Science Foundation under Grant No.~DMR-1944539.

\section{Appendix}
\label{sec:appendix}
We investigated the performance of the \emph{three-move nonreversible Berretti-Sokal algorithm} in $4D$ and $5D$. We did not find it to be efficient, when compared to the \emph{two-move nonreversible Berretti-Sokal algorithm}. The detailed findings are in the table. 
\begin{table}[H]
    \centering
    \begin{tabular}{|c|cc|}
   \hline
    \multicolumn{3}{|c|}{dimension $d = 4$}\\
   \hline
    system size $n$&$x=6/35$&$x=3/35$\\
    \hline
   $ 25$ &$0.714 \pm 0.069$ & $2.970 \pm 0.356$
   \\
   $ 51 $ &$1.081 \pm 0.050$ & $2.216 \pm 0.229$ \\
   $ 75 $ &$0.994 \pm 0.033$ & $2.812 \pm 0.658$ \\
   $ 101 $ &$0.945 \pm 0.028$ & $2.349 \pm 0.190$\\
   \hline
   \hline
   \multicolumn{3}{|c|}{dimension $d = 5$} \\ 
   \hline
    system size $n$ &$x=1/5$&$x=1/10$\\
    \hline
    $ 21$ &$0.920 \pm 0.002$ & $4.214 \pm 1.108$\\
    $ 25$ &$0.961 \pm 0.001$ & $4.451 \pm 0.571$\\
    $ 31$ &$0.992 \pm 0.002$ & $4.992 \pm 0.696$\\
    $ 35$ &$0.995 \pm 0.002$ & $3.261 \pm 0.513$\\
    \hline
\end{tabular}    
    \caption{The ratio of integrated autocorrelation times of the \emph{three-move nonreversible Berretti-Sokal algorithm}, $\tau$, and the \emph{two-move nonreversible Berretti-Sokal algorithm}, $\tau_0$, for $4D$ and $5D$ systems as a function of the linear system size $n$, the SAW unit length weight $x$. The ratio about 1 for ($x = 6/35$, $d = 4$) and ($x = 1/5$, $d = 5$), however for $(x = 3/35, d = 4)$ and $(x = 1/10, d = 5)$ it is above unity, which indicates that two-mode nonreversible Berretti-Sokal algorithm is more efficient there.}
    \label{tab:3move-NBS}
\end{table}

\bibliography{VuceljaBib.bib}

\begin{thebibliography}{10}

\bibitem{old}
Hu, H., Chen, X., and Deng, Y.: {Irreversible Markov chain Monte Carlo
  algorithm for self-avoiding walk}.
\newblock \emph{Frontiers of Physics} 12, 120503 (2016)

\bibitem{Havlin_1982}
Havlin, S. and Ben-Avraham, D.: New approach to self-avoiding walks as a
  critical phenomenon.
\newblock \emph{Journal of Physics A: Mathematical and General} 15, L321--L328
  (1982)

\bibitem{PhysRevA.26.1728}
Havlin, S. and Ben-Avraham, D.: Theoretical and numerical study of fractal
  dimensionality in self-avoiding walks.
\newblock \emph{Phys Rev A} 26, 1728--1734 (1982)

\bibitem{2002math......4277L}
{Lawler}, G.~F., {Schramm}, O., and {Werner}, W.: {On the scaling limit of
  planar self-avoiding walk}.
\newblock \emph{arXiv Mathematics e-prints} math/0204277 (2002)

\bibitem{53Flory}
{Flory}, P.: \emph{{Principles of Polymer Chemisty}}.
\newblock Cornell University Press (1953)

\bibitem{mc1}
Metropolis, N. and Ulam, S.: {{T}he {M}onte {C}arlo method}.
\newblock \emph{J Am Stat Assoc} 44, 335--341 (1949)

\bibitem{mcmc}
van Rensburg, E. J.~J.: {Monte Carlo methods for the self-avoiding walk}.
\newblock \emph{Journal of Physics A: Mathematical and Theoretical} 42, 323001
  (2009)

\bibitem{poly1}
{de Carvalho}, C., Caracciolo, S., and Frohlich, J.: Polymers and $g|\phi|^4$
  theory in four dimensions.
\newblock \emph{Nuclear Physics B} 215, 209--248 (1983)

\bibitem{poly2}
Duplantier, B.: Polymer Network of fixed topology: renormalization, exact
  critical exponent $\ensuremath{\gamma}$ in two dimensions, and
  $d=4\ensuremath{-}\ensuremath{\epsilon}$.
\newblock \emph{Phys Rev Lett} 57, 941--944 (1986)

\bibitem{ising1}
Zhou, Z., Grimm, J., Fang, S., Deng, Y., and Garoni, T.~M.: Random-Length
  Random Walks and Finite-Size Scaling in High Dimensions.
\newblock \emph{Phys Rev Lett} 121, 185701 (2018)

\bibitem{ising2}
Zhou, Z., Grimm, J., Deng, Y., and Garoni, T.~M.: Random-length Random Walks
  and Finite-size Scaling on high-dimensional hypercubic lattices I: Periodic
  Boundary Conditions (2020)

\bibitem{nspin}
Fang, S., Deng, Y., and Zhou, Z.: Logarithmic Finite-Size Scaling of the
  Self-avoiding Walk at Four Dimensions (2021)

\bibitem{97Sokal}
Sokal, A.: Monte Carlo Methods in Statistical Mechanics: Foundations and New
  Algorithms.
\newblock In DeWitt-Morette, C., Cartier, P., and Folacci, A., eds.,
  \emph{Functional Integration}, vol. 361 of \emph{NATO ASI Series}, 131--192.
  Springer US, New York, NY, USA (1997).
\newblock ISBN 978-1-4899-0321-1

\bibitem{NewmanMC}
Newman, M. E.~J. and Barkema, G.~T.: \emph{{Monte Carlo Methods in Statistical
  Mechanics}}.
\newblock Clarendon Press (1999)

\bibitem{pivot}
Madras, N. and Sokal, A.~D.: {The pivot algorithm: A highly efficient Monte
  Carlo method for the self-avoiding walk}.
\newblock \emph{Journal of Statistical Physics} 50, 109--186 (1988)

\bibitem{bsalgorithm}
Berretti, A. and Sokal, A.~D.: New {Monte Carlo} method for the self-avoiding
  walk.
\newblock \emph{Journal of Statistical Physics} 40, 483--531 (1985)

\bibitem{BFACF}
van Rensburg, E. J.~J. and Whittington, S.~G.: The {BFACF} algorithm and
  knotted polygons.
\newblock \emph{Journal of Physics A: Mathematical and General} 24, 5553--5567
  (1991)

\bibitem{DHN97}
Diaconis, P., Holmes, S., and Neal, R.~M.: Analysis of a non-reversible Markov
  chain sampler.
\newblock \emph{Technical Report BU-1385-M}  (1997)

\bibitem{CLP00}
Chen, F., Lovasz, L., and Pak, I.: {Lifting Markov Chains to Speed up Mixing}.
\newblock \emph{Proceedings of the ACM symposium on Theory of Computing}
  275--281 (1999)

\bibitem{turitsyn2011irreversible}
Turitsyn, K.~S., Chertkov, M., and Vucelja, M.: Irreversible {Monte Carlo}
  algorithms for efficient sampling.
\newblock \emph{Physica D Nonlinear Phenomena} 240, 410--414 (2011)

\bibitem{LiftingVucelja}
Vucelja, M.: {Lifting--A nonreversible Markov chain Monte Carlo algorithm}.
\newblock \emph{American Journal of Physics} 84, 958--968 (2016)

\bibitem{2013SakaiHukushima}
{Sakai}, Y. and {Hukushima}, K.: {Dynamics of One-Dimensional Ising Model
  without Detailed Balance Condition}.
\newblock \emph{Journal of the Physical Society of Japan} 82, 064003--1--8
  (2013)

\bibitem{bierkens2017}
Bierkens, J. and Roberts, G.: A piecewise deterministic scaling limit of lifted
  Metropolis--Hastings in the Curie--Weiss model.
\newblock \emph{Ann Appl Probab} 27, 846--882 (2017)

\bibitem{joris}
Bierkens, J.: {Non-reversible Metropolis-Hastings}.
\newblock \emph{Statistics and Computing} 26, 1213--1228 (2016)

\bibitem{17KapferKrauth}
Kapfer, S.~C. and Krauth, W.: Irreversible Local Markov Chains with Rapid
  Convergence towards Equilibrium.
\newblock \emph{Phys Rev Lett} 119, 240603 (2017)

\bibitem{atm0}
van Rensburg, E. J.~J. and Rechnitzer, A.: Atmospheres of polygons and knotted
  polygons.
\newblock \emph{Journal of Physics A: Mathematical and Theoretical} 41, 105002
  (2008)

\bibitem{atm1}
Rechnitzer, A. and van Rensburg, E. J.~J.: Canonical {Monte Carlo}
  determination of the connective constant of self-avoiding walks.
\newblock \emph{Journal of Physics A: Mathematical and General} 35, L605--L612
  (2002)

\bibitem{atm2}
Rechnitzer, A. and van Rensburg, E. J.~J.: {Generalized atmospheric Rosenbluth
  methods} ({GARM}).
\newblock \emph{Journal of Physics A: Mathematical and Theoretical} 41, 442002
  (2008)

\bibitem{endpointatm}
van Rensburg, E. J.~J. and Rechnitzer, A.: Generalized atmospheric sampling of
  self-avoiding walks.
\newblock \emph{Journal of Physics A: Mathematical and Theoretical} 42, 335001
  (2009)

\bibitem{2009Levinbook}
Levin, D.~A., Peres, Y., and Wilmer, E.~L.: \emph{Markov Chains and Mixing
  Times}.
\newblock American Mathematical Society, Providence, RI, USA (2009)

\bibitem{MRRTT53}
Metropolis, N., Rosenbluth, A., Rosenbluth, M., Teller, A., and Teller, E.:
  Equations of State Calculations by Fast Computing Machines.
\newblock \emph{J of Chem Phys} 21, 1087--1092 (1953)

\bibitem{H70}
Hastings, W.~K.: Monte Carlo sampling methods using Markov chains and their
  applications.
\newblock \emph{Biometrika} 57, 97--109 (1970)

\bibitem{autotime}
{Goodman}, J. and {Weare}, J.: {Ensemble samplers with affine invariance}.
\newblock \emph{Communications in Applied Mathematics and Computational
  Science} 5, 65--80 (2010)

\bibitem{PERM1}
Hsu, H.-P. and Grassberger, P.: Polymers confined between two parallel plane
  walls.
\newblock \emph{The Journal of Chemical Physics} 120, 2034--2041 (2004)

\bibitem{PERM2}
Owczarek, A.~L. and Prellberg, T.: Scaling of self-avoiding walks in high
  dimensions.
\newblock \emph{Journal of Physics A: Mathematical and General} 34, 5773--5780
  (2001)

\bibitem{parrelPERM}
Campbell, S. and van Rensburg, E. J.~J.: Parallel {PERM}.
\newblock \emph{Journal of Physics A: Mathematical and Theoretical} 53, 265005
  (2020)

\bibitem{NIPS2017_7099}
Guyon, I., Luxburg, U.~V., Bengio, S., Wallach, H., Fergus, R., Vishwanathan,
  S., and Garnett, R., eds.: \emph{{A-NICE-MC}: Adversarial Training for
  {MCMC}}. Curran Associates, Inc. (2017)

\end{thebibliography}

\end{document}